\documentclass[aps,11pt,nofootinbib, tightenlines]{revtex4-2}
\pdfoutput=1
\usepackage{graphicx}
\usepackage{float}
\usepackage[margin=1in]{geometry}
\usepackage{amsmath}
\usepackage{amssymb}
\usepackage[utf8]{inputenc}
\usepackage[all]{nowidow}

\newcommand{\PBH}{\mathrm{PBH}}
\newcommand{\CDM}{\mathrm{CDM}}
\newcommand{\diffd}{\mathrm{d}} 
\newcommand{\ddv}[2]{\frac{\diffd #1}{\diffd #2}} 

\usepackage[linktocpage=true,colorlinks=true,linkcolor=blue,citecolor=blue,urlcolor=blue]{hyperref}
\hypersetup{pdftitle={Smooth vs instant inflationary transitions: steepest growth re-examined and primordial black holes},pdfauthor={4 authors}}

\begin{document}

\author{Philippa S.~Cole$^1$}
\email{p.s.cole@uva.nl}
\author{Andrew D.~Gow$^{2,3}$}
\email{andrew.gow@port.ac.uk}
\author{Christian T.~Byrnes$^3$}
\email{C.Byrnes@sussex.ac.uk}
\author{Subodh P. Patil$^4$}
\email{patil@lorentz.leidenuniv.nl}

\affiliation{\\1) GRAPPA Institute, \\\mbox{University of Amsterdam, 1098 XH Amsterdam, The Netherlands}\\}

\affiliation{\\2) Institute of Cosmology and Gravitation, \mbox{University of Portsmouth, Portsmouth, PO1 3FX, United Kingdom}\\}

\affiliation{\\3) Department of Physics and Astronomy,\\University of Sussex, Brighton BN1 9QH, United Kingdom\\}

\affiliation{\\4) Instituut-Lorentz for Theoretical Physics,\\Leiden University, 2333 CA Leiden, The Netherlands\\}

\date{03/05/2024}

\title{Smooth vs instant inflationary transitions: steepest growth re-examined and primordial black holes}

\begin{abstract}
Primordial black holes (PBHs) can be produced by a range of mechanisms in the early universe. A particular formation channel that connects PBHs with inflationary phenomenology invokes enhanced primordial curvature perturbations at small scales. In this paper, we examine how rapidly the background can transition between different values of the parameters of the Hubble hierarchy in the context of single-clock inflation, which must ultimately derive from a consistent derivative expansion for the background inflaton field. We discuss artefacts associated with instant or very rapid transitions, and consider the impact on the steepest power spectrum growth and the formation of PBHs. In particular, we highlight the robustness of the $k^4$ steepest growth previously found for single-field inflation with conservatively smoothed transitions and limits on how much the amplitude of the power spectrum can grow, and demonstrate that the PBH mass distribution is sensitive to the artefacts, which go away when the transitions are smoothed. We also show that the mass distribution is relatively insensitive to the steepness of the growth and subsequent decay of the power spectrum, depending primarily on the peak amplitude and the presence of any plateaus that last more than an e-fold. The shape of the power spectrum can of course be constrained by other tracers, and so understanding the physical limitations on its shape remains a pertinent question.
\end{abstract}

\maketitle

\newpage
\tableofcontents

\section{Introduction}

Primordial black holes (PBHs) could teach us a lot about the initial conditions of the universe. Large overdensities must have been left over at the end of inflation for PBHs to have formed during radiation domination, when pressure forces strongly oppose gravitational collapse, see \cite{Sasaki:2018_Primordial,Carr:2020_Constraints,Carr:2020_Primordial,Green:2020_Primordial} for recent reviews. Assuming that PBH formation occurs via the direct collapse of these large amplitude perturbations shortly after they re-enter the horizon post-inflation, and that the perturbations follow a Gaussian distribution, the primordial power spectrum needs to be around seven orders of magnitude larger than the amplitude observed on large scales \cite{Gow:2021_ACPS}. This raises questions of how quickly the power spectrum can grow by this amount and how the shape of a peak in the power spectrum could be observed. 

Since the publication of \cite{Byrnes:2019_Steepest}, where it was observed that the fastest the primordial power spectrum can grow under certain assumptions in single-field inflation is $k^4$, examples have been found of inflationary potentials that enable the primordial power spectrum to grow slightly faster in single-field inflation \cite{Carrilho:2019_Dissecting,Ozsoy:2020_Slope,Ragavendra:2020sop,Tasinato:2020vdk}, and substantially faster in multi-field inflation \cite{Palma:2020_Seeding,Fumagalli:2020_Turning,Braglia:2020taf}. Many models that are able to produce enough growth for PBH production employ sharp transitions in the inflationary parameters, e.g. \cite{Pi_2023,Cai_2022,Inomata_2021,Ivanov:1994pa}, notably Starobinsky's piece-wise potential~\cite{Starobinsky:1992ts}, although we note that the original paper actually invokes a smoothed transition rather than an instant one. While it is an attractive quality of these models that it is often possible to approximate the dynamics analytically, the presence of the instant or rapid transitions can lead to unphysical features in the resulting power spectrum and primordial black hole mass function whether they are calculated analytically or numerically. We study various properties of enhancements to the power spectrum with the aim of understanding which of them are artefacts of unrealistic instant or rapid transitions in the underlying model, and which are physical. We show that features such as oscillations in the power spectrum that arise from unrealistic instant transitions in the evolution of the inflationary parameters disappear if the transitions instead occur over the duration of approximately an e-fold. Instantaneous in the context of inflationary cosmology should mean an order one fraction of an e-fold (a distinction that for instance, the authors of \cite{Felder:1998vq, Felder:1999pv} were keenly aware of in their definition of `instantaneous' in the context of pre-heating). We also show that the appearance of an over-optimistic growth in amplitude of the power spectrum can be produced, which goes away when transitions are realistically smoothed. In a follow-up paper we will study the fine-tuning associated with building inflationary models capable of generating PBHs.

We use these results to further demonstrate the importance of using a realistically smoothed primordial power spectrum to calculate the resulting PBH abundance. For a power-law growth in the power spectrum, the resultant distribution or abundance of PBHs is largely insensitive to changes in the slope of the power spectrum peak. In particular, we find that \textit{there is essentially no dependence on the steepness of the power spectrum once it is steeper than $k^2$}, not even compared to the unrealistic (but frequently studied) delta-function power spectrum. However, we show that there can be large differences to the resulting distribution of PBHs between the unsmoothed and smoothed power spectra that we calculate in section \ref{sec:matching}, due to the fact that power spectrum peaks can broaden substantially and even lead to multi-modal PBH mass distributions. We note that regardless of the PBH mass function, observational constraints on primordial density fluctuations can be placed, for example via observations of the stochastic gravitational wave background or CMB spectral distortions (depending on the relevant comoving scales). These may be sensitive to changes in the shape of the power spectrum and can provide extra information about the underlying mechanism that generated the enhanced power. Hence, the shape of an enhancement to the power spectrum is an important factor if we are to constrain it with other probes, and yet more reason to ensure a realistic form for the power spectrum is used when calculating constraints.

For simplicity, we do not include the impact of non-Gaussianity of the perturbations in this paper, despite PBH formation taking place in the tail of the probability distribution which implies that PBH formation is sensitive to non-Gaussianity. Many studies of non-Gaussianity in ultra-slow-roll (USR) inflation have been made, see e.g.~\cite{Namjoo:2012aa,Cai:2018dkf,Ozsoy:2021pws,Passaglia_2019}, as well as on PBH formation in this and related scenarios \cite{Atal:2018neu,Atal:2019erb,Ragavendra:2020sop,Davies:2021loj,Figueroa:2021zah,Inomata:2021tpx}, but the impact this has on PBH formation remains uncertain. Issues include whether the (order unity) non-Gaussianity has a physical impact or not (related to its value in Conformal Fermi Coordinates) \cite{Bravo:2017gct,Bravo:2017wyw,Matarrese:2020why,Suyama:2021adn} and the impact of quantum diffusion during USR inflation \cite{Pattison:2017mbe,Pattison:2019hef,Ezquiaga:2019ftu,Pattison:2021oen,Biagetti:2021eep,Ballesteros:2020sre}.

In section \ref{sec:matching} we demonstrate the important impact of smoothing the inflationary potential, going beyond the instantaneous transitions assumed for analytic simplicity. In section \ref{sec:PBH} we study the impact of the shape of the power spectrum peak on PBH mass distributions, including the impact of the smoothed transitions from section \ref{sec:matching}. We conclude in section \ref{sec:conclusions}.

\section{Steepest growth: artefacts from instant transitions}
\label{sec:matching}

\begin{figure}[H]
\centering
\includegraphics[width=0.8\textwidth]{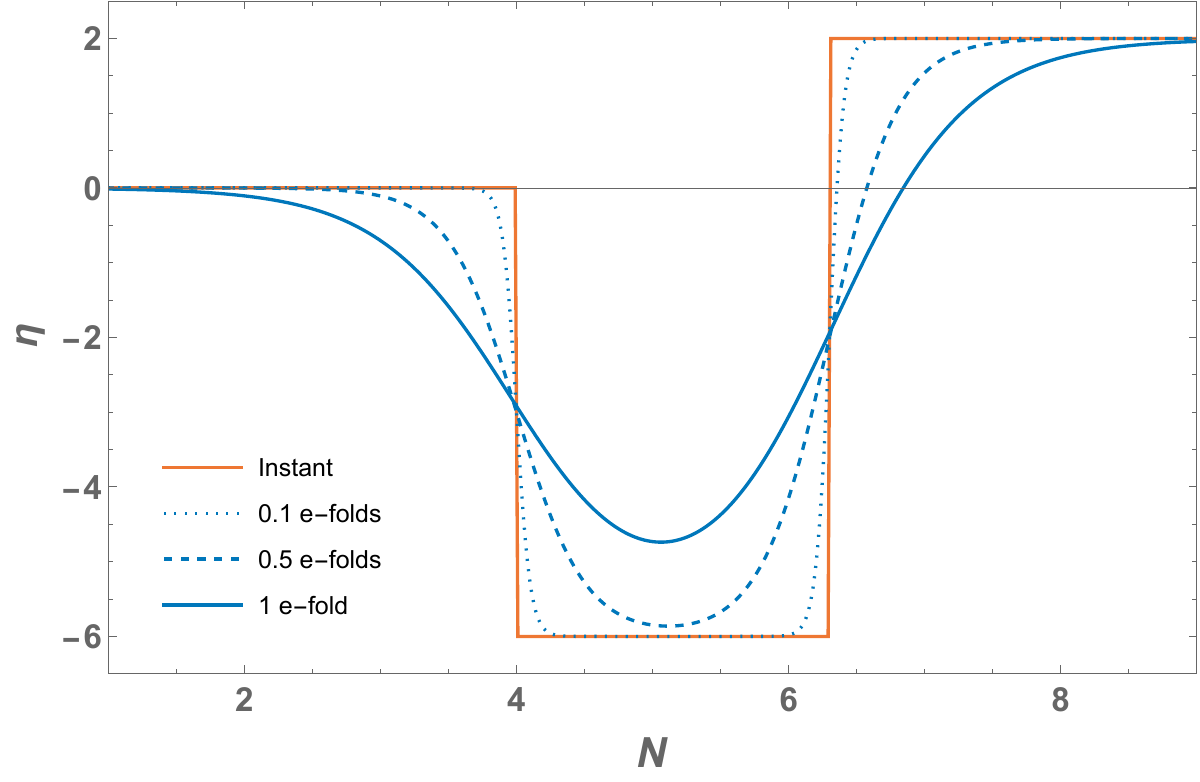}
\caption{Trajectories of $\eta$ as a function of e-folds. Instant transitions are shown in orange, and varying durations of smoothing are shown in dotted (0.1 e-folds), dashed (0.5 e-folds) and solid (1 e-fold) blue.}
\label{fig:eta} 
\end{figure}

Perhaps the strongest motivation for considering a significant growth in the power spectrum is the production of primordial black holes, which can contribute to early- and late-universe physics in a number of ways, most notably as a dark matter candidate. Many PBH-motivated models exhibit artefacts in the resultant power spectrum such as rapid oscillations in $k$-space, usually around the peak\footnote{Note that this scenario is distinct from the perturbative oscillations generated on CMB scales in models where the power spectrum amplitude remains close to constant, see e.g.~\cite{Stewart_2002,Arroja_2011}.}. These are consequences of instant or rapid transitions in the underlying model, which can allow for the power spectrum to be calculated analytically by stitching together the solutions for the curvature perturbation that are each valid for a constant value of the second slow-roll parameter $\eta$. The slow-roll parameters are defined as $\epsilon=-\dot{H}/H^2$ and $\eta=\dot{\epsilon}/\epsilon H$, and it is the latter which tracks the acceleration of the inflaton and determines the slope of the resulting power spectrum. For a summary of the literature, see for instance \cite{Atal:2018neu}. The solutions (and their derivatives) are `matched' at transition times between phases of constant $\eta$, resulting in an evolution of $\eta$ that has instant transitions between constant values. The details of such matching calculations are laid out in \cite{Byrnes:2019_Steepest}. See for example figure~\ref{fig:eta}, where the solid orange line shows the $\eta$ evolution which corresponds to instant transitions between slow-roll, followed by ultra-slow-roll $(\eta=-6)$ and then a period of $\eta=2$ (which induces a decay in the power spectrum on small scales). Whilst one can imagine that stitching together a large number of these solutions for constant $\eta$ will eventually approach the limit of a smoothly evolving function of $\eta$, the instant transitions are unrealistic for large changes in the value of $\eta$ at the transition times. This is because $\eta$ and $\epsilon$ are obtained from perturbing around a time dependent background derived from an underlying effective action for the background inflaton, and are thus subject to the usual requirements of deriving from a controlled derivative expansion.

Instantaneous transitions correspond to arbitrarily sharp changes in certain parameters of the effective action (i.e.~the potential and its derivatives in the context of canonical inflation), which are unphysical, because they do not correspond to a controlled derivative expansion over the field range of interest. Concretely, although one might try to engineer an arbitrarily sharp transition at the classical level, this will not be possible at the quantum corrected (i.e. renormalisation group improved) level\footnote{This can be straightforwardly seen through approximating any sharp feature in a potential by a Taylor series in the field operator. Even if one requires a proliferation of higher order monomials  to approximate the feature within some radius of convergence (for which the beta functions are known in closed form to all orders at one loop for a single scalar \cite{Garny:2008mig}), the coefficients in the Taylor expansion rapidly become greater than unity. This results in quantum corrections so large that the effective potential will bear no resemblance to the classical potential that ordinarily serves as a good approximation in the case where instead of a feature, the potential has an approximate shift symmetry over the field range of interest.}. In addition, one must take care to distinguish between being able to engineer arbitrarily sharp transitions in coordinate time, and in terms of e-folds. Large gradients source expansion, and transitions between different values of the parameters of the Hubble hierarchy always end up costing an order one fraction of an e-fold even if one were to suspend disbelief and try to arrange for them to occur arbitrarily fast in coordinate time (this is discussed in appendix C2 of \cite{Byrnes:2019_Steepest} in the context of how quickly inflation can end, but the argument can be generalised). We should therefore instead consider smoothed transitions.  We model the smoothing phenomenologically by multiplying each phase of constant $\eta$ by
\begin{equation}
s(N,\nu)=\frac{1}{2} \left[\tanh{\left(\frac{N - N_i}{\nu}\right)} - \tanh{\left(\frac{N - N_j}{\nu}\right)} \right],
\end{equation}
where $N$ denotes e-folds, $N_i$ and $N_j$ are the e-fold values at the transition times in and out of the phase of constant $\eta$, and $\nu$ is the duration of the smoothing. The effect on the evolution of $\eta$ is demonstrated by the dotted, dashed and solid blue lines of figure~\ref{fig:eta} which correspond to $\nu=(0.1,0.5,1)$ e-fold transition durations respectively. There are 2.3 e-folds of ultra-slow-roll in the case of instant transitions.

\begin{figure}[H]
\centering
\includegraphics[width=0.8\textwidth]{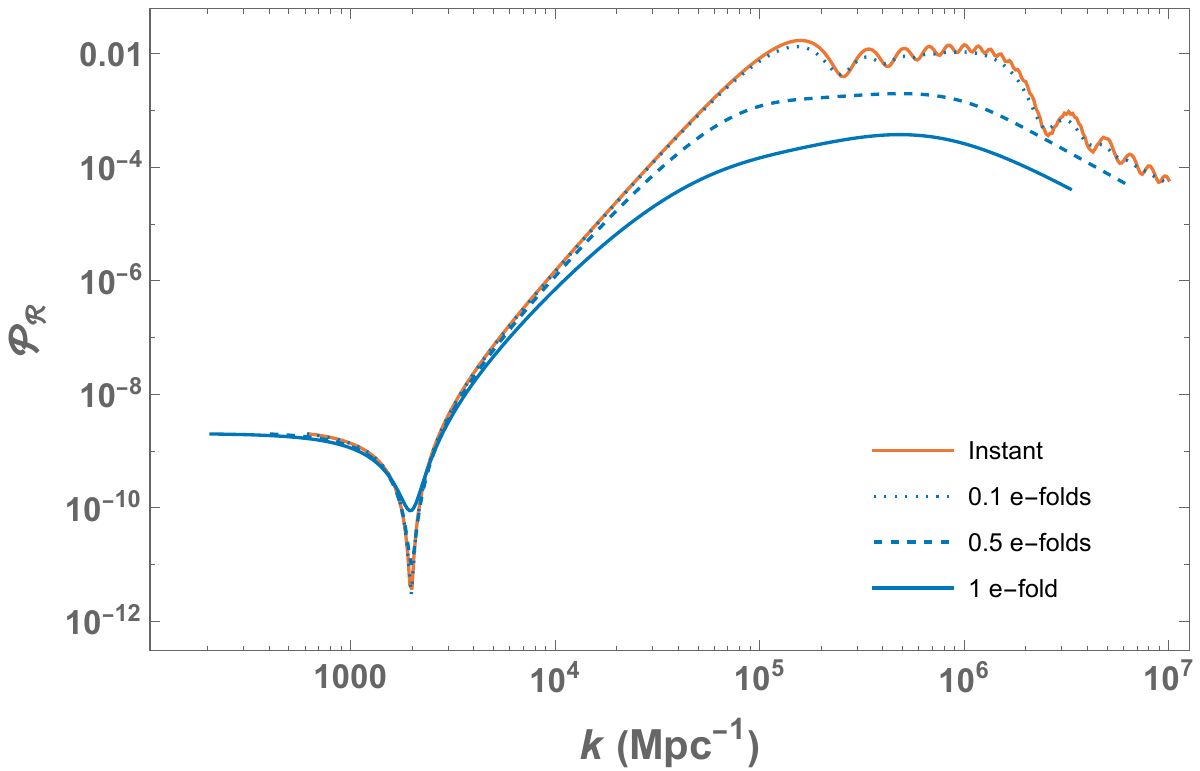}
\caption{Power spectra for instant and smoothed $\eta$ trajectories. The analytic matching solution with instant transitions is shown in orange, and the smoothed numerical results with varying `durations' of smoothing are shown in dotted (0.1 e-folds), dashed (0.5 e-folds) and solid (1 e-fold) blue.}
\label{fig:smooth}
\end{figure}

The corresponding power spectra are shown in figure~\ref{fig:smooth}. The spectrum calculated from matching instantaneously between constant phases of $\eta$ (solid orange line) exhibits oscillations around the peak of the spectrum. Given that these alter the peak amplitude, they could have a noticeable effect on the PBH mass distribution and abundance, and they would certainly impact observations which directly constrain the scalar perturbations. When we instead smooth the transitions, the oscillations disappear. The dotted, dashed and solid blue lines in figure~\ref{fig:smooth} correspond to 0.1, 0.5 and 1 e-fold durations of smoothing of the $\eta$ evolution respectively. For 0.1 e-folds of smoothing, the oscillations survive, but for 0.5 and 1 e-folds, they disappear entirely. Furthermore, the peak amplitude decreases too, which will have an exponentially large effect on the abundance of PBHs produced. However, this is in part due to the fact that when smoothing the transitions, the duration of non-slow-roll is decreased, so one would expect less growth in the power spectrum.

\begin{figure}[H]
\centering
\includegraphics[width=0.8\textwidth]{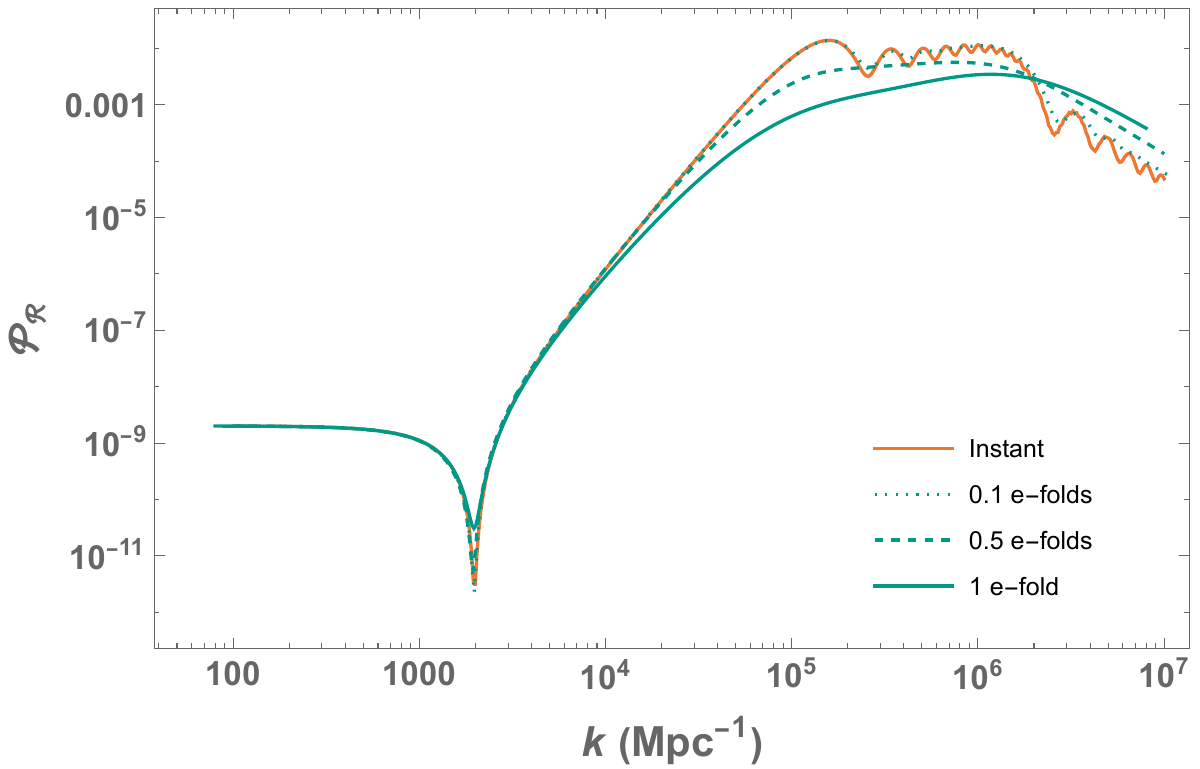}
\caption{Power spectra for instant and smoothed $\eta$ trajectories. The analytic matching solution with instant transitions is shown in orange, and the smoothed numerical results with varying `durations' of smoothing are shown in dotted (0.1 e-folds), dashed (0.5 e-folds) and solid (1 e-fold) teal. The integrated contribution of negative $\eta$ is kept fixed for all 4 lines.}
\label{fig:smooth_equal_eta}
\end{figure}

To provide a fairer comparison between the unsmoothed and smoothed cases, we can approximately normalise the growth of the power spectrum. The total power spectrum growth is connected to the total integrated amount of $\eta < 0$ evolution (i.e.~$\int^{N_2}_{N_1}\eta(N)\,\diffd N$ where $\eta(N)<0$ for $N_1<N<N_2$). Therefore, we carry out the same smoothing procedure as for figure~\ref{fig:smooth}, but change the duration of the ultra-slow-roll phase in each case to keep this integrated value constant. Figure~\ref{fig:smooth_equal_eta} shows the same unsmoothed power spectrum as figure~\ref{fig:smooth} in orange, and the smoothed and normalised power spectra in teal, with the same smoothing durations as figure~\ref{fig:smooth}. We now see that the oscillations disappear, and the peak height is still decreased but less significantly.

It is important to note here that the process of smoothing the instant transitions causes the peak of the power spectrum to decrease and shift to larger values of $k$, and also flattens the slope in the vicinity of the peak. These changes will have a significant effect on the PBH mass distribution, since this is only sensitive to the region close to the peak. We will address this effect in section \ref{sec:PBH} (see figure~\ref{fig:smooth_psiPlot}).

\begin{figure}[H]
\centering
\includegraphics[width=0.8\textwidth]{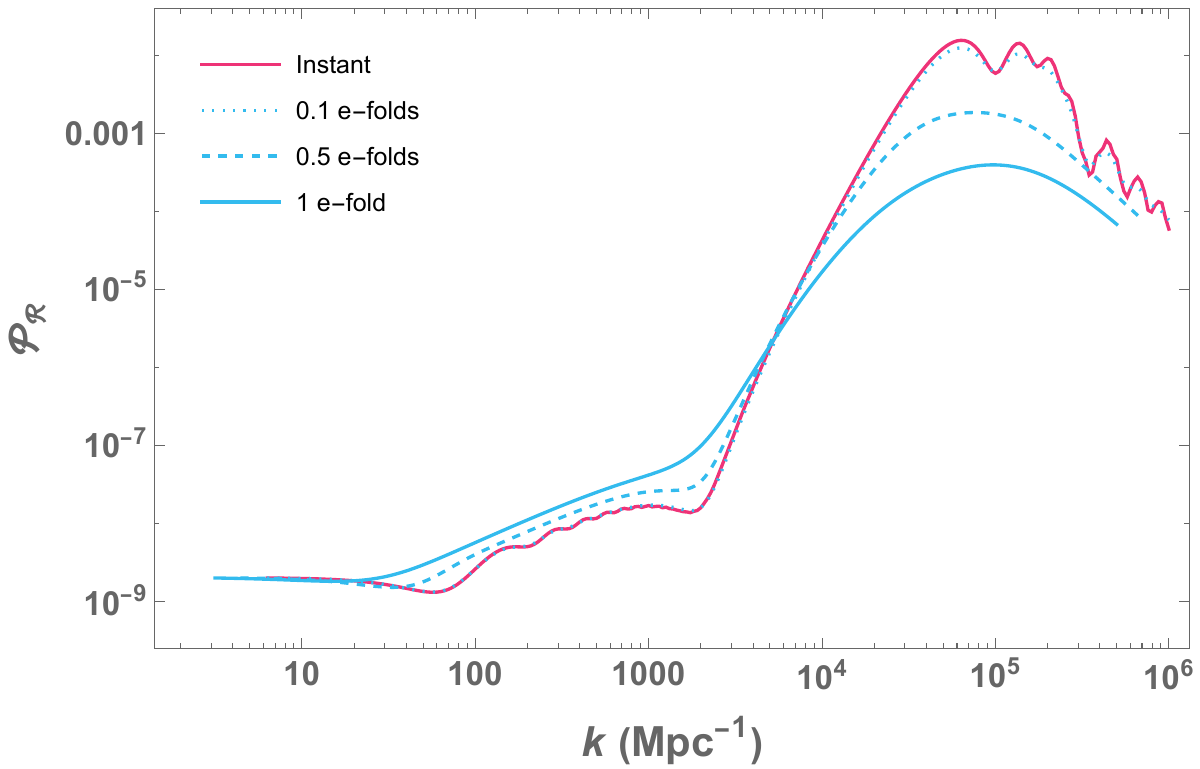}
\caption{Power spectra including an extra phase of 6 e-folds of $\eta=-1$ evolution in between slow-roll and ultra-slow-roll (1.3 e-folds thereof). The analytic matching solution with instant transitions is shown in magenta, and the smoothed numerical results with varying `durations' of smoothing are shown in dotted (0.1 e-folds), dashed (0.5 e-folds) and solid (1 e-fold) light blue.}
\label{fig:qmslope}
\end{figure}

Finally we also demonstrate that models which exhibit super-$k^4$ growth with instant transitions between constant phases of $\eta$ suffer from similar oscillation, slope and amplitude depletion when instant transitions are smoothed. For example, it was shown in \cite{Carrilho:2019_Dissecting} that an extra phase of $\eta=-1$ in between phases of slow-roll and ultra-slow-roll produces an enhancement of the $k^4$ slope in the power spectrum proportional to $k^5(\log{k})^2$. By smoothing the transitions and limiting the amplitude increase in the primordial power spectrum to no more than seven orders of magnitude so as to comply with both large scale CMB constraints and the envelope of PBH non-detection constraints, we find that the super-$k^4$ growth is effectively lost, as shown in figure~\ref{fig:qmslope}. This happens for exactly the same reason as the $k^4$ growth depletion shown in figures~\ref{fig:smooth} and \ref{fig:smooth_equal_eta}. However, because just one matching is required between slow-roll and ultra-slow-roll, some $k^4$ growth does survive. In cases where other constant phases of $\eta$ are required with a strong sensitivity on the duration of that phase, for example the extra phase of $\eta=-1$ in this model, the impact of super-$k^4$ power spectrum growth is overruled by the need to smooth transitions\footnote{We note in closing that our arguments only apply to \textit{sustained} growth, i.e., growth which can be achieved for multiple e-folds. There are transients where the growth index is evidently much higher (such as in the immediate vicinity of the dip before the sustained growth). One can imagine positing classical inflationary backgrounds where transients of a specific nature can be arranged to occur regularly enough so that the growth can have periods of super $k^4$ growth that last more than a few e-folds, as was shown in \cite{Ozsoy:2020_Slope, Ragavendra:2020sop}, although whether these persist at the quantum corrected level is to be verified.}.

We will now review the effect of the total slope of the power spectrum, as well as whether it plateaus or decays, on the PBH mass distribution. We will then go on to show that spurious features close to the peak of the power spectrum, which we have shown in this section should be smoothed, can lead to unrealistically large effects on the PBH mass distribution which will not manifest from a reasonable power spectrum.

\section{PBH mass distribution: effects of growth and decay of the power spectrum}
\label{sec:PBH}
A key observable for the study of PBHs is their mass distribution, which can be sensitive to the details of the power spectrum of the perturbations from which they form. However, it is known that the effects of critical collapse become significant for narrow power spectra \cite{Niemeyer:1998_Critical-collapse,Yokoyama:1998_Critical-collapse,Kuhnel:2016_Critical-collapse}, and it has recently been shown that if the power spectrum peak is sufficiently narrow, the resulting PBH mass distribution is described entirely by the critical collapse physics and becomes insensitive to the detailed shape of the power spectrum \cite{Gow:2020cou}. It is therefore sensible to ask the question of whether any growth steeper than $k^4$ is distinguishable from $k^4$ growth when considering the PBH mass distribution.

We calculate the PBH mass distribution using the formalism described in \cite{Gow:2021_ACPS}. There is considerable discussion about which method to use when calculating the PBH mass distribution, with many recent papers utilising a peaks theory method. It was shown in \cite{Gow:2021_ACPS} that the mass distribution shape is mostly independent of the calculation method once normalised to a fixed $f_\PBH$. As there is no community consensus about the correct way to apply peaks theory~\cite{Yoo:2018kvb,Yoo:2019pma,Germani:2019zez,Young:2020_Peaks}, we choose to use the Press--Schechter method for simplicity. Additionally, we neglect the effect of the non-linear relationship between $\zeta$ and $\delta$ \cite{Kawasaki:2019mbl,Young:2019yug,DeLuca:2019qsy,Kalaja:2019_From}. This is expected to have some impact on the mass distribution shape, but will not affect the conclusions presented in this section. For these choices the total PBH density is given by\footnote{We neglect the impact of PBH accretion and evaporation in this paper.}
\begin{align}
\Omega_\PBH &= 2\int \diffd(\ln R)\ \frac{R_\text{eq}}{R}\int_{\delta_{R,c}}^{\infty} \diffd\delta_R\ \frac{m}{M_H} P(\delta_R), \label{eq:Omega-PBH}
\end{align}
where $\delta_R$ is a version of the density contrast smoothed on a scale $R$, for which we use the modified Gaussian window function from \cite{Gow:2021_ACPS}. The ratio of masses is given by the critical collapse equation
\begin{align}
m &= K M_H (\delta_R - \delta_{R,c})^\gamma, \label{eq:Critical-collapse}
\end{align}
where we use the parameters $K = 10$, $\gamma = 0.36$, and $\delta_{R,c} = 0.25$. $P(\delta_R)$ is assumed to be Gaussian, with a variance related to the curvature perturbation power spectrum through
\begin{align}
\sigma^2(R) &= \int_{0}^{\infty} \frac{\diffd k}{k}\ \frac{16}{81}(kR)^4 W^2(kR) \mathcal{P}_\zeta(k),
\end{align}
where $W(kR)$ is the window function. The mass distribution is then given by
\begin{align}
\psi(m) &= \frac{1}{\Omega_\PBH}\ddv{\Omega_\PBH}{m},
\end{align}
which is normalised to unity over the PBH mass $m$. In this expression, equation~\eqref{eq:Critical-collapse} must be inverted to write the inner integral in equation~\eqref{eq:Omega-PBH} in terms of $m$, and the horizon mass is written in terms of $R$. This mass distribution is related to another commonly used definition for the PBH mass function, $f(m)$, by
\begin{align}
\psi(m) &= \frac{1}{f_\PBH}\frac{f(m)}{m},
\end{align}
where $f(m)$ satisfies the normalisation condition
\begin{align}
\int f(m)\ \diffd \ln(m) &= f_\PBH,
\end{align}
with $f_\PBH = \Omega_\PBH/\Omega_\CDM$ the fraction of dark matter in the form of PBHs.

For the shape of the primordial power spectrum, we consider a symmetric peak with a power-law growth and decay of a constant slope, and for simplicity we begin by assuming that the growth and decay rates of the slope are equal,
\begin{align}
\mathcal{P}_\mathcal{R}(k) &= A\begin{cases}
\left(\frac{k}{k_p}\right)^n & k \leqslant k_p \\
\left(\frac{k}{k_p}\right)^{-n} & k > k_p
\end{cases}, \label{eq:Pzeta_sym}
\end{align}
where $k_p$ is the peak scale and the amplitude $A$ is fixed to produce a PBH abundance $f_\PBH = 10^{-3}$.

\begin{figure}[H]
\centering
\includegraphics[width=0.8\textwidth]{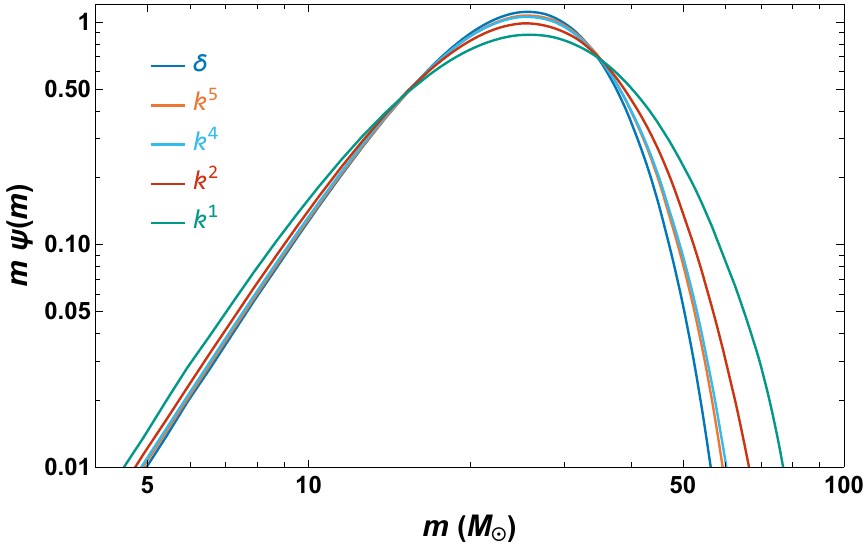}
\caption{PBH mass distribution from the symmetric power-law power spectrum in equation~\eqref{eq:Pzeta_sym}. The peak positions have been normalised to the same value for ease of comparison. For sufficiently steep peaks, the mass distribution becomes virtually indistinguishable.}
\label{fig:PBH_sym}
\end{figure}

In figure~\ref{fig:PBH_sym} we show the PBH mass distributions calculated from these symmetric power spectrum peaks, for a range of slopes. It is clear that a slope of $k^4$ is already steep enough that the PBH mass distribution is extremely close to the distribution resulting from a delta function in the power spectrum. This means that, by extension, the PBH mass distribution resulting from any power spectrum slope steeper than $k^4$ will be even less distinguishable from the mass function produced by a delta function power spectrum.

\begin{figure}[H]
\centering
\includegraphics[width=0.5\textwidth]{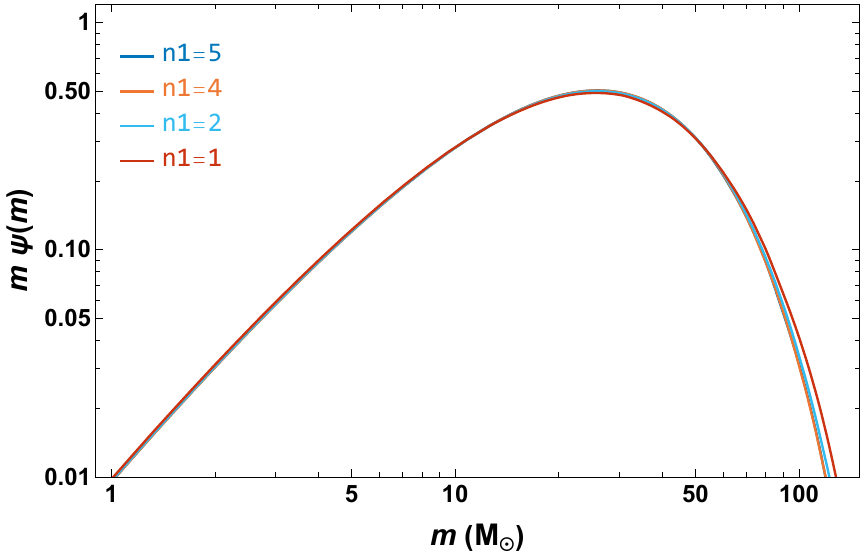}\includegraphics[width=0.5\textwidth]{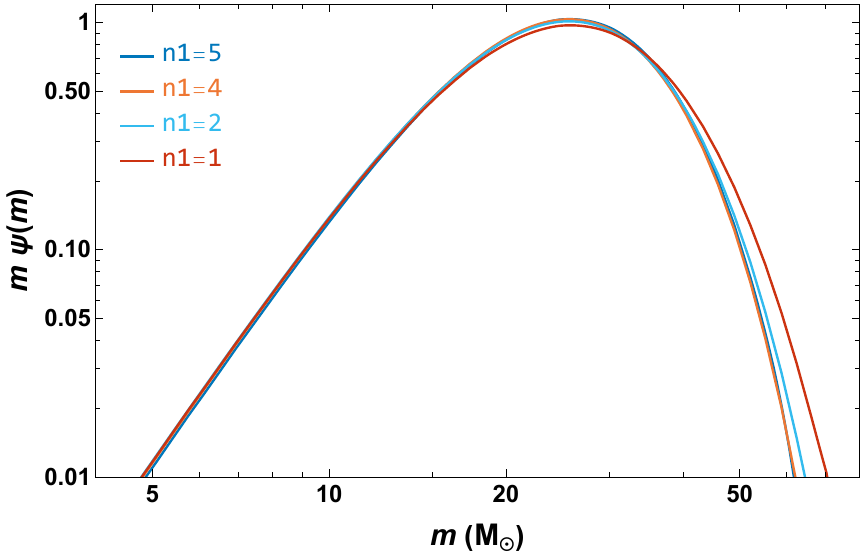}
\caption{PBH mass distribution from the asymmetric power-law power spectrum in equation~\eqref{eq:Pzeta_asym} with $n_2 = 0.15$ (left) and 2.7 (right). The peak positions have been normalised to the same value for ease of comparison. As with the case of a symmetric power spectrum, the mass distribution becomes virtually indistinguishable for sufficiently fast growth.}
\label{fig:PBH_asym}
\end{figure}

However, the assumption that the power spectrum will fall off with the same slope as its rising edge is not normally valid in practice. For example, in inflationary models with an inflection point, the growth is usually $k^4$ whilst the decrease in all known cases is less steep and related to the value of the second slow-roll parameter after the inflection point. Therefore, it is necessary to consider the case of an asymmetric power spectrum peak, given by
\begin{align}
\mathcal{P}_\mathcal{R}(k) &= A\begin{cases}
\left(\frac{k}{k_p}\right)^{n_1} & k \leqslant k_p \\
\left(\frac{k}{k_p}\right)^{-n_2} & k > k_p
\end{cases}, \label{eq:Pzeta_asym}
\end{align}
with the amplitude again normalised to give $f_\PBH = 10^{-3}$. For the rising edge, we will consider the same range of indices as for the symmetric case. For the falling slope, we choose the bounding values given in \cite{Atal:2018neu} for the selection of models they considered, $0.15 \leqslant n_2 \leqslant 2.7$. In figure~\ref{fig:PBH_asym} we show the mass distributions for these two cases. It is clear that the conclusion stated above for symmetric peaks also holds for asymmetric power spectra.

Another possibility is that the power spectrum does not decay immediately, but remains enhanced over some range of $k$ values. This is the case for the example of a simple USR phase ($\eta = 0\to-6\to0$), in which the negative $\eta$ period grows the power spectrum, after which it remains at its enhanced value. This simple case is not possible to realise in many inflationary models, since the end of inflation normally requires $\epsilon=1$, and so a positive $\eta$ phase must exist to drive $\epsilon$ back to larger values, which will also cause the power spectrum to reduce. However, it may be possible to maintain an enhanced power spectrum for a small number of e-folds before it decays again, or to end inflation by an alternative mechanism to the growth of $\epsilon$, such as a waterfall field in hybrid inflation. Therefore, we consider the case of a power spectrum that grows and decays as $k^4$, but with a plateau at the top,
\begin{align}
\mathcal{P}_\mathcal{R}(k) &= A\begin{cases}
\left(\frac{k}{k_l}\right)^4 & k \leqslant k_l \\
1 & k_l < k \leqslant k_u \\
\left(\frac{k}{k_u}\right)^{-4} & k > k_u
\end{cases},
\end{align}
where $k_u$ is defined in terms of the width of the plateau in log-space,
\begin{align}
k_u &= k_le^w
\end{align}
and the amplitude is fixed such that $f_\PBH=10^{-3}$.

\begin{figure}[H]
\centering
\includegraphics[width=0.5\textwidth]{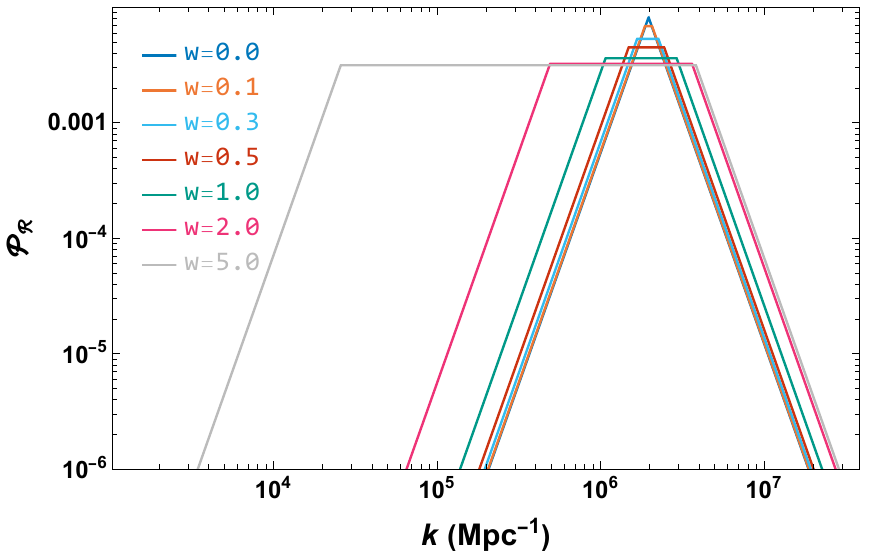}\includegraphics[width=0.5\textwidth]{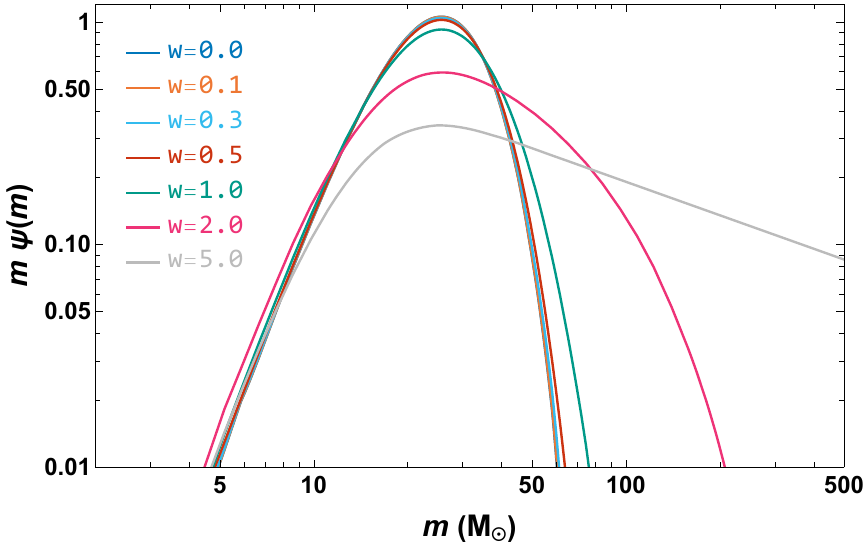}
\caption{Plateau power spectra (left) and corresponding PBH mass distributions (right) for a range of plateau widths $w$. The mass distribution peak positions have been aligned for ease of comparison.}
\label{fig:PBH_plateau}
\end{figure}

In figure~\ref{fig:PBH_plateau} we show a range of plateau power spectra with different widths $w$, ranging from no plateau to one that lasts for five e-folds (where e-folds $N$ relate to comoving scales $k$ as $N\propto\log{k}$), and the corresponding PBH mass distributions. The power spectrum positions are chosen such that the mass distributions peak at the same mass. We can see that for plateau widths smaller than $\sim$ one e-fold (i.e.~$k_u/k_l\simeq2.7$), the mass distribution is virtually unchanged. For broader plateaus, an enhancement to the high-mass tail is observed.

Why is the PBH mass function so insensitive to substantial changes in the shape of the power spectrum peak? There are two key reasons. Firstly, the density perturbations are due to the curvature power spectrum smoothed with a window function, which smears out features on the scale of the PBH (see e.g.~figure~5 of \cite{Byrnes:2019_Steepest} to see the visual impact of the smoothing, or \cite{Kalaja:2019_From}). While there has been some discussion as to the correct choice of window function, it now seems clear that the effect on the PBH mass distribution is sufficiently small to be indistinguishable with current data \cite{Ando:2018_Primordial,Young:2020_Criterion,Gow:2021_ACPS}. The second reason is that the critical collapse phenomenon implies a minimum width to the PBH mass distribution even if the power spectrum is a delta function (meaning that it only has perturbations on a single scale). This ensures that a monochromatic mass distribution is impossible, and a lognormal shape is frequently used instead. However, for sufficiently narrow power spectra, the shape is controlled by the critical collapse effect, leading to a non-lognormal distribution, and an insensitivity to the power spectrum shape \cite{Niemeyer:1998_Critical-collapse,Yokoyama:1998_Critical-collapse,Gow:2020cou}. The combination of these two effects means that features in the power spectrum with a width corresponding to less than about one e-fold are unlikely to be visible to any realistic future measurement of the PBH mass function.

\begin{figure}[H]
\centering
\includegraphics[width=0.8\textwidth]{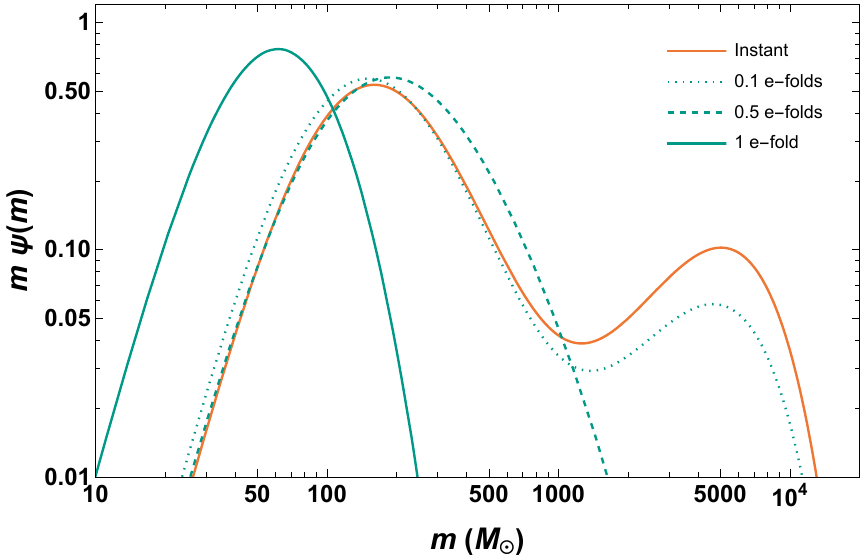}
\caption{PBH mass distribution corresponding to the power spectra shown in figure~\ref{fig:smooth_equal_eta}. The mass distribution corresponding to the analytic matching solution with instant transitions is shown in orange, and those corresponding to the smoothed numerical results with varying `durations' of smoothing are shown in dotted (0.1 e-folds), dashed (0.5 e-folds) and solid (1 e-fold) teal.}
\label{fig:smooth_psiPlot}
\end{figure}

It is also important to stress that the PBH formation rate is only sensitive to the power spectrum amplitude very close to the peak, because the collapse rate into PBHs is exponentially sensitive to the amplitude. Hence, the PBH mass function can (at best) only be a function of the power spectrum shape close to the peak. An example of this is shown by calculating the PBH mass distributions corresponding to the power spectra shown in figure~\ref{fig:smooth_equal_eta}. While the unsmoothed power spectrum in figure~\ref{fig:smooth_equal_eta} exhibits oscillations, these are removed by the window function. However, the unsmoothed power spectrum still gives a mass distribution with a unique double peak structure (solid orange line in figure~\ref{fig:smooth_psiPlot}), corresponding to the primary maximum and the broader oscillatory second peak in the power spectrum. However, it can be seen that smoothing the transitions quickly suppresses the second peak, before removing it entirely. For 0.5 e-folds of smoothing (dashed teal line in figure~\ref{fig:smooth_psiPlot}), the power spectrum has a plateau, producing a broad mass distribution, and for one e-fold (solid teal line in figure~\ref{fig:smooth_psiPlot}), the low-$k$ peak in the power spectrum no longer contributes to the PBH mass distribution, leaving just a single narrow peak shifted to lower masses. Therefore, we can see that any search for a bimodal mass distribution, see e.g.~\cite{Cai:2018tuh,Carr:2018poi,Hall:2020daa,Zheng:2021vda,Inomata:2018cht}, is not motivated by ultra-slow-roll inflation once realistic transitions are assumed.

Alternative probes will be needed to constrain or detect the power spectrum on amplitudes between the $10^{-9}$ value seen on the largest scales and the $\sim 10^{-2}$ required for PBH formation. On scales at most a few orders of magnitude smaller than the smallest scales detectable with the CMB or LSS (relevant for the formation of PBHs massive enough to be detected with LIGO--Virgo--KAGRA), one can use spectral distortions of the CMB \cite{Zeldovich:1969ff, Sunyaev:1970er, Hu:1992dc, Chluba:2011hw, Khatri:2012tw} or, in a more model dependent way, ultracompact minihaloes \cite{Bringmann:2012_UCMH,Gosenca:2017_UCMH,Sten_Delos:2018_UCMH,Sten_Delos:2018_UCMH-density,Nakama:2019htb} or gravitational lensing of small scale structures \cite{Karami:2018qrl,Lee:2020wfn,Wagner-Carena:2022mrn}. One weakness of these alternative constraints is that the $\mu$-distortions measure an integrated value of the small scale power, while the other two techniques only apply for certain dark matter models, and we note that a mixed dark matter model with a combination of WIMPs and PBHs is incompatible over a large range of parameter space \cite{Lacki:2010_Primordial,Adamek:2019_WIMPs,Carr:2020mqm,Boudaud:2021irr}.

An important method to detect the primordial perturbations on small scales is via the primordial stochastic gravitational wave background, generated by the large amplitude scalar perturbations sourcing second-order tensor perturbations, see \cite{Yuan:2021qgz,Domenech:2021ztg} for reviews. There is an approximate one-to-one relation between the PBH mass and the peak of the gravitational wave frequency. At the time of writing the stochastic gravitational wave background constraint is only competitive with the PBH constraint on the power spectrum amplitude on scales around one parsec, with the corresponding frequency probed by pulsar timing arrays (PTAs), see e.g.~\cite{NANOGrav:2020bcs}. Interestingly, this constraint is almost exactly equal to the PBH constraint on the mass range in which LIGO--Virgo--KAGRA can detect black holes, so if that experiment does detect (or has already detected) PBHs then there should be a corresponding signal detectable by PTAs. In the future, other experiments including SKA, LISA and the Einstein Telescope will be able to probe the stochastic gravitational wave background with sufficient sensitivity over such a large range of scales that a non-detection by all of these experiments would potentially be able to rule out the existence of almost any (non-evaporated) PBHs assuming that they form from the direct collapse of Gaussian distributed density perturbations. For the current and forecast constraint plots see figures~5 and 6 of~\cite{Gow:2021_ACPS} (see also \cite{Wang:2019kaf}). However, if any of these experiments detect a primordial stochastic gravitational wave background, then measurements of its frequency dependence over a wide range of frequencies could provide a measurement of the shape of the power spectrum peak, e.g.~\cite{Pi:2020otn,Ragavendra:2020sop,Dalianis:2021_Exploring,Domenech:2021ztg}.

\section{Conclusions}
\label{sec:conclusions}

We have shown that transitions between different periods of inflation have a large effect on the resulting primordial power spectrum. Models with instantaneous transitions between periods with different growth rates of $\eta$ are often studied both analytically (possible exactly because of the instant transitions) and numerically. However, we caution that instant transitions are unphysical and that realistic power spectra (with only numerically tractable solutions), based on physically realisable smooth transitions, have somewhat different behaviour. The ``steepest'' growth of $k^4$ in most ultra-slow-roll scenarios persists \cite{Byrnes:2019_Steepest}, but the oscillations after the peak are damped into something close to a plateau and the amplitude of the peak is also reduced. In at least one scenario capable of getting a steeper than $k^4$ growth~\cite{Carrilho:2019_Dissecting}, we have shown that this steeper growth is erased upon smoothing the transitions to reasonable durations.

We have also studied the implications of the shape of a large peak in the primordial power spectrum on the primordial black hole mass function. We have shown that changes in the steepness of the peak, provided the steepness is significantly greater than $k$ before the peak, or $k^{-1}$ after the peak, have a negligible impact on the PBH mass function. As a corollary, we can also see that growth rates faster than the ``common'' limit of $k^4$ will not have any measurable impact on the PBH mass function, and even assuming an unrealistic delta-function power spectrum is adequate for this purpose. Although the conclusion that the PBH mass function has a minimum width (due to critical collapse and the impact of the smoothing when going from the curvature perturbation to the density contrast) is not new, our detailed study builds upon previous investigations. We have studied peaks with an equal power law growth and decay, peaks with an unequal growth and decay rate (as motivated by single-field models of ultra-slow-roll inflation) and also peaks including a plateau of variable width. In the case of a peak with a plateau, we have shown that plateau widths which correspond to scales exiting over less than about one e-fold ($N\propto\log{k}$) have a minimal impact on the PBH mass function, but a plateau width corresponding to scales exiting over two e-folds has a significant impact on boosting the large mass tail.

In light of this in-depth review of the limited effect of different power spectrum slopes on the mass function, we go on to show that power spectra from realistically smoothed inflationary transitions produce significantly different PBH mass distributions than those from instant transitions, thereby demonstrating the importance of calculating PBH mass distributions from initial conditions with smooth trajectories for the inflationary slow-roll parameters.

\section*{Acknowledgements}
The authors thank Sam Young for useful discussions. PC acknowledges support from the Institute of Physics at the University of Amsterdam.  AG acknowledges support from the Science and Technology Facilities Council [grant numbers ST/S000550/1 and ST/W001225/1]. CB acknowledges support from the Science and Technology Facilities Council [grant number ST/T000473/1]. For the purpose of open access, the authors have applied a Creative Commons Attribution (CC BY) licence to any Author Accepted Manuscript version arising. No new raw data were generated or analysed in support of this research.

\bibliographystyle{JHEP-edit} 
\bibliography{PBH_model_building_paper_1}{}

\end{document}